\documentstyle[twoside,fleqn,espcrc2,epsf]{article}

\newcommand{\tr}{{\rm Tr}}
\newcommand{\re}{{\rm Re}}

\newcommand{\retr}{\re\,\tr}
\newcommand{\be}{\begin{equation}}
\newcommand{\ee}{\end{equation}}
\newcommand{\bea}{\begin{eqnarray}}
\newcommand{\eea}{\end{eqnarray}}

\newcommand{\AmS}{{\protect\the\textfont2
  A\kern-.1667em\lower.5ex\hbox{M}\kern-.125emS}}

\title{SU(2) Colour Fields around
Static Sources}

\author{\underline{G.S.~Bali},
C.~Schlichter and K.~Schilling\thanks{Work supported
        by DFG grant Schi 257/1-4 and EC project SC1*-CT91-0642.}
       \vskip\baselineskip
        Physics Department,
        University of Wuppertal, 42097 Wuppertal, Germany}

\begin{document}

\begin{abstract}
First results of an ongoing high statistics
study of the colour flux distribution
around static quark sources in $SU(2)$ gauge theory are presented.
The flux tube profiles and widths have been investigated for several
quark separations at $\beta=2.5$ and $\beta=2.74$. The results are
tested against Michael's sum rules.
\end{abstract}

\maketitle

\section{INTRODUCTION}
Recently, high statistics studies of the static $q\bar q$ potential
in $SU(2)$ and $SU(3)$ gauge theories have been
performed~\cite{bali}. A linear long range potential with
the universal subleading $-\pi/(12R)$ correction
(for quark separations $Ra>0.5$~fm), as predicted by the
string picture~\cite{luescher}, is observed.
Moreover, the gap between the ground
state potential and the first excited state
potential for large $R$ is
consistent with the string value of
$\pi/R$~\cite{michpara}.
The hope for
a deeper understanding of the underlying dynamics of the
confinement mechanism is one of the motivations for the present
investigation. Effective models like a bosonic string
model~\cite{luescher}, based on a Nielsen-Olesen szenario
of magnetic confinement~\cite{nielsen}
can be tested.

Measurements of colour flux distributions are extremely difficult
because the relevant correlators of pure glue
quantities exhibit large statistical fluctuations.
For this reason, previous studies (e.g.~Refs.~\cite{sommer,prev})
have only been
qualitative. In view of the great progress achieved recently in
hardware and lattice techniques
(improved operators, noise reduction, updating
algorithm), it appears to be of considerable interest to perform really
reliable quantitative studies of field distributions. I will report on
first results of an ongoing project with this aim.

\section{SIMULATION}
We study lattice volumes of $16^4$ at $\beta=2.5$
($a\approx 0.085$~fm) and $32^4$ at $\beta=2.74$ ($a\approx 0.041$~fm).
Thus, the physical volumes are $(1.3$~fm$)^4$ in either case. The scale
has been extracted from the string tension value
$\sqrt{\sigma}=440$~MeV. For the updating a hybrid of heatbath and
overrelaxation has been used (details in Ref.~\cite{schlicht}).
Measurements of the relevant operators have been taken
every 100 sweeps.  At the two $\beta$ values, 8640 and 670 such
statistically independent measurements have been taken, respectively.
For safety, the data was binned into blocks of {\em five}
prior to the analysis.

The chromoelectric (and -magnetic) field distributions in
presence of two static quarks have been measured for various quark
separations $r=Ra$.
The quarks are
generated at time $t=0$, and annihilated at time $t=Ta$.
This is realized by use of Wilson loops $W(R,T)$.
A variant of APE smearing~\cite{APE}
has been applied to the spatial parts of the Wilson
loops in order to increase the ground state overlaps. All overlaps
have been found to be larger than $95\%$ in the potential analysis.

\begin{figure}[htb]
\leavevmode
\vskip -1.5cm
\epsfxsize=200pt
\epsfbox[55 130 655 530]{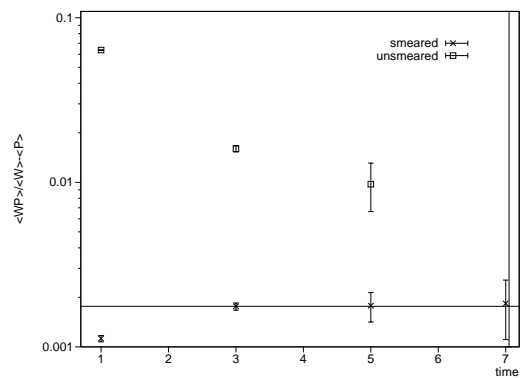}
\caption{Comparison between smeared and unsmeared operators for
$\varepsilon({\bf 0})$ at $\beta=2.5, R=4$.}
\label{fig1}
\end{figure}
The Maxwell field strength tensor is given by
${F^a_{\mu\nu}}^2=2\tr
{\cal F}^2_{\mu\nu}\approx 2\beta/a^4S_{\mu\nu}$
with $S_{\mu\nu}=1-\frac{1}{N}
\retr\left(U_{\mu\nu}\right)$ being the plaquette
action. Due to its locality, $S_{\mu\nu}$ undergoes
a multiplicative renormalization which, in the mean field
approximation, is cancelled out in the
combination $\beta S_{\mu\nu}$.

We measure the difference between this operator in presence of the
$q\bar q$ pair, separated by a distance $r$, and its vacuum
expectation for various times $T$:
\begin{equation}
P^{(R,T)}_{\mu\nu}({\bf n})=\frac{\langle W(R,T)S_{\mu\nu}({\bf
n},T/2)\rangle}{\langle W(R,T)\rangle}-\langle S_{\mu\nu}\rangle.
\end{equation}
The (squared) field components are given by
\bea
\label{fields}
{E_{(r)}}_i^2({\bf n}a)&=&\frac{2\beta}{a^4}P^{(R,T)}_{i4}({\bf n})\quad,\\
{B_{(r)}}_i^2({\bf n}a)
&=&\frac{2\beta}{a^4}\epsilon_{ijk}P^{(R,T)}_{jk}({\bf n})
\eea
for $T$ large.
{}From these fields, the energy, $\varepsilon({\bf x})$, and action
densities, $\sigma({\bf x})$, are calculated:
\be
\left.\begin{array}{c}\varepsilon_{(r)}({\bf x})\\
                \sigma_{(r)}({\bf x})\end{array}\right\}
=\frac{1}{2}\left({\bf E}_{(r)}^2({\bf x})
\pm{\bf B}_{(r)}^2({\bf x})\right)\quad.
\ee
Since $B_i^2\leq 0$ in
the chosen Minkowski notation, the energy densities are very small
and, thus, extremely difficult to measure.
In the actual simulation, various electric/magnetic plaquettes have
been averaged, in order to obtain an operator that is symmetric
around the lattice site $({\bf n},T/2)$.

Early investigations have failed to identify the asymptotic plateau
(Eqs.~(2,3)). This situation is greatly improved by smearing techniques,
as can be seen from Fig.~\ref{fig1} where a comparison
of $\varepsilon_{(4a)}({\bf 0})$ (at $\beta=2.5$)
from smeared and unsmeared Wilson loops is made.
Also, the statistical errors are substantially reduced (note
the logarithmic scale).
The coordinate system is chosen such that the quark source resides at the
spatial coordinate $(0,0,R/2)$, and the antiquark at $(0,0,-R/2)$. The
fields have been measured at all points ${\bf n}$ with
$n_3$ being varied along the whole lattice axis, while
the transverse distance $n_{\perp}=(n_1,n_2)$ is taken along the
directions $(1,0)$, and $(1,1)$ up to $|n_{\perp}|=6$.

\begin{figure}[htb]
\leavevmode
\vskip -.9cm
\epsfxsize=200pt
\epsfbox[55 130 655 530]{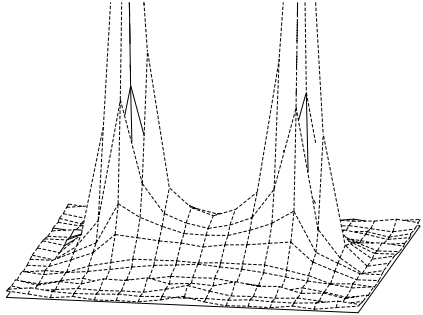}
\caption{The action density $\sigma({\bf n})$ at $\beta=2.5$,
$r=8a\approx 0.7$~fm.\vskip -.5cm}
\label{fig2}
\end{figure}

\section{RESULTS}
In Fig.~\ref{fig2}, the measured action density
(at $\beta=2.5$, $r=8a\approx 0.7$~fm) is displayed.
The resolution of our lattice allows us to
see smooth physical structures without any recourse to fancy
interpolations. Note that the mesh is not equidistant in the
perpendicular direction
because the off-axis separations $n_{\perp}\propto (1,1)$ are included.

It is instructive to investigate the
field components parallel ($E^2_{\parallel}=E^2_z$)
and perpendicular ($E^2_{\perp}=E_x^2+E_y^2$) to the flux tube.
Around the middle of the confining string (${\bf n}=0$)
$B_{\parallel}^2<B_{\perp}^2\stackrel{<}{\sim}
E_{\perp}^2\ll E_{\parallel}^2$ is observed.
Nice scaling behaviour of the
colourfields is found between the two $\beta$-values.
Within statistical accuracy, continuum rotational invariance is
restored.

The energy and action distributions in the central perpendicular plane
($n_3=0$) have been investigated.
Gaussian fits $\varepsilon(x_{\perp},0)=c^2\exp(-x_{\perp}^2/b^2)$
and dipole fits $\varepsilon(x_{\perp},0)=c^4/(x_{\perp}^2+b^2)^3$
have been performed. In addition,
we have numerically integrated the densities
within a circle with the cut-off radius $x_{\mbox{\scriptsize max}}$
\be
I_{\varepsilon}(f,r)=\int_{x_{\perp}^2\leq x^2_{\mbox{\tiny max}}}
\!\!\!\!\!\!\!\!\!\!d^2\!x_{\perp}\,\varepsilon_{(r)}(x_{\perp},0)\,f(x_{\perp})\quad.
\ee
\(x_{\mbox{\scriptsize max}}\)
has been varied in order to identify a plateau.
The width is defined by
$\rho_{\varepsilon}^2(r)=I_{\varepsilon}(x_{\perp}^2,r)/I_{\varepsilon}(1,r)$.
For the above fit functions one obtains $\rho_{\varepsilon}^2=b^2$.
For $r<0.4$~fm the dipole fits yield better $\chi^2$
values. For larger $q\bar q$ separations
both $\chi^2$ turn out to be acceptable but the Gaussian results
are closer to the numerically integrated values. The
same holds true for $\rho_{\sigma}$.
The electric field ${\bf E}^2(x_{\perp},0)$ is expected to
have a Gaussian shape within the confining string~\cite{luescher}.

\begin{figure}[htb]
\leavevmode
\vskip -1.3cm
\epsfxsize=200pt
\epsfbox[55 130 655 530]{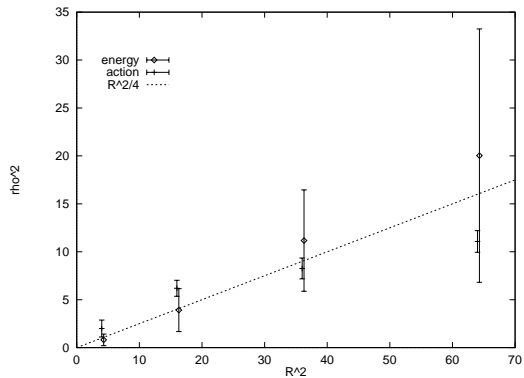}
\caption{Widths of the energy and action flux tubes
$\rho^2_{\epsilon/\sigma}$ (numerically integrated)
as a function of $R^2$
in lattice units.\vskip -.5cm}
\label{fig3}
\end{figure}
$\rho^2_{\varepsilon}$ and $\rho^2_{\sigma}$ are
plotted against $R^2$ in Fig.~\ref{fig3}.
Both, the action, and the energy density widths
increase with $R$. In the Coulomb-dominated region of the potential
($R<0.4$~fm) a dipole
behaviour $\rho^2_{E^2}(R)=R^2/4$ is expected. This
straight line is plotted together with the data.
For large separations we expect a logarithmic dependence
from the string picture~\cite{luescher}.
The actual lattice volumes
are still too small to investigate this effect.

\begin{figure}[htb]
\leavevmode
\vskip -1.3cm
\epsfxsize=200pt
\epsfbox[55 130 655 530]{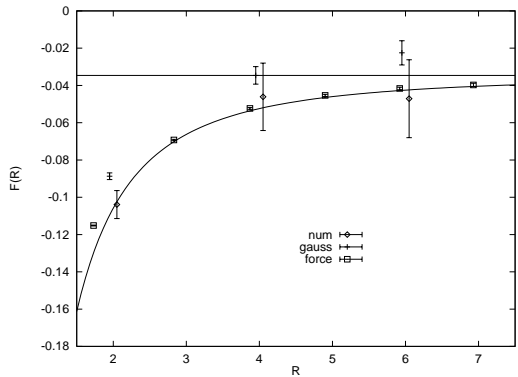}
\caption{Comparison between the (Gaussian and numerically) integrated
perpendicular energy density and the interquark force (Eq.~6).}
\label{fig4}
\end{figure}
The integrated energy density should equal the potential between the
quarks up to an additive self energy~\cite{mich}
(and the multiplicative field
renormalization which is approximately {\em one}
for the lattice field operators chosen).
Under the assumption that on increasing the $q\bar q$ separation,
the additional binding energy is completely localized
near the center of the two the sources (which
is expected as soon as an effectively one dimensional string has
been formed), a differential form of this sum rule can be derived:
\be
\label{force}
F(R)= -a^2\!\!\int\!d^2x_{\perp}\,\varepsilon_{(Ra)}(x_{\perp},0)\quad,
\ee
where $F(R)=-\partial V(R)/\partial R$ is the interquark force (in
lattice units).
This relation is tested in
Fig.~\ref{fig4}. Agreement
is found even for $R=2$, a distance, where the
Coulomb contribution to the potential clearly dominates.
This is reflected in an extreme stability of the shapes of the
field distributions near the sources
against increasing the quark separation.

Till now, we have only realized physical quark
separations smaller than $0.7$~fm,
where a one dimensional string is not yet really formed.
Thus, small differences between the fields at $n_3=0$
and $n_3=1$ are still visible.
Work on a lattice of linear extent $L\approx 2.7$~fm is
in progress~\cite{schlicht}.

\end{document}